\documentclass[11pt,a4paper,nologo]{ETHpaper}
\usepackage{graphicx, amsmath, amssymb}
\usepackage[square,numbers,sort&compress]{natbib}
\usepackage{cleveref}
\usepackage{bm}
\usepackage{booktabs,dcolumn}
\newcommand{\abs}[1]{\lvert#1\rvert}
\newcommand{\G}{\mathcal{I}}

\title{Multiplex Network Regression: \\How do relations drive interactions?}

\renewcommand{\thefootnote}{\fnsymbol{footnote}}

\author{Giona Casiraghi\footnote{E-mail: gcasiraghi@ethz.ch.}}
\address{Chair of Systems Design, ETH Z\"urich, Weinbergstrasse 56/58, 8092 Z\"urich, Switzerland.}
\authoralternative{G. Casiraghi}
\reference{(Submitted for publication: 10 July 2020)}
\www{\url{http://www.sg.ethz.ch}}

\begin{document}
\makeframing
\maketitle

\renewcommand{\thefootnote}{\arabic{footnote}}
\addtocounter{footnote}{-4}

\begin{abstract}
\noindent We introduce a statistical regression model to investigate the impact of dyadic relations on complex networks generated from observed repeated interactions.
It is based on generalised hypergeometric ensembles (gHypEG), a class of statistical network ensembles developed recently to deal with multi-edge graph and count data.
We represent different types of known relations between system elements by weighted graphs, separated in the different layers of a multiplex network.
With our method, we can regress the influence of each relational layer, the explanatory variables, on the interaction counts, the dependent variables.
Moreover, we can quantify the statistical significance of the relations as explanatory variables for the observed interactions. 
To demonstrate the power of our approach, we investigate an example based on empirical data.

\end{abstract}

\section{Introduction}
\label{sec:introduction}

In the study of real-world complex systems, we often deal with datasets of \emph{observed repeated interactions} between individuals.
These datasets are used to generate networks where system's elements are represented by vertices and interactions by edges.
We ask whether these interactions are random events or whether they are driven by existing relations between system's elements.
To answer this question, we propose a statistical model to regress \emph{relations}, which we identify as \emph{covariate variables}, on a network created from interactions, which we will refer to as our \emph{dependent variables}.

In general, a regression model explains dependent variables as a function of some covariates, accounting for random effects.
Here, we assume that the observed interactions are driven by different relations, possibly masked by \emph{combinatorial effects}.
With combinatorial effects, we mean that elements that interact more, in general, are also more likely to interact with each other, even if they have no relations.
This problem is well known in network theory, where it is referred to as \emph{degree-correction} (see e.g., \citep{Peixoto2014a,Newman2015,Karrer2011}).
For example, the fact that two individuals have contact very often can be explained by multiple reasons.
They may interact because they are friends, because work together, or simply because they are very active, and hence have high chances to meet.
Therefore, to have a full understanding of a system, we have to disentangle relations from combinatorial effects.

Datasets of interactions are ubiquitous across disciplines.
Examples of these are recorded contacts between individuals (e.g., SocioPatterns~\citep{Stehle2011,Mastrandrea2015}, Reality Mining~\citep{Eagle2006}), mutualistic interactions between species in ecology~\citep{Memmott1999,Dicks2002}, economical transactions between countries and firms~\citep{Garas2010,Chakraborti2003}, and collaborations between firms \citep{tomasello2016rise}.
In these cases, researchers are interested in learning whether the observed interactions are driven by relations between the elements of the system.
They ask whether friendship plays a role in the contacts between students~\citep{Mastrandrea2015}, whether \emph{homophily} drives interactions within social and political networks~\citep{brandenberger2018trading}, i.e., whether individuals sharing similar characteristics are more likely to interact~\citep{Mcpherson2001}, or whether collaborations between companies are driven by geographical distance or industrial sector similarity~\citep{tomasello2016rise}.

There exist different approaches addressing the problem of \emph{quantifying} the interdependence between observed edges and dyadic relations in networks.
This problem, however, is exacerbated by the fact that the dyadic relations represented in complex networks are not independent of one another.
There is a broad literature on modelling relational data to account for some of these properties that can arguably be traced to the Social Relations Model of~\cite{Warner1979}.
Because of the non-independency of dyadic relations, ordinary least squares regression models are inappropriate to analyse network data~\citep{Krackhardt1988}.
To partially overcome their limits, \citep{Krackhardt1988} introduced a regression method based on the quadratic assignment procedure developed by~\citep{Hubert1976}.
An alternative approach to address the problem of the non-independence of edges is that taken in latent space models~\citep{Hoff2002}.
There, although the model still assumes the probability of edges to be independent in the sampling process, the dependence is accounted for in the latent space constructed from the data.
Other statistical methods commonly used in the analysis of social networks are based on exponential random graph models (ERGMs) or their extensions (see, e.g.,~\citep{Snijders1996,Snijders2010,Snijders2011,Snijders1995}).
Although being effective under specific conditions, all these methods have been developed for \emph{unweighted} graphs.
This means that they are not optimal for datasets which contain repeated interactions, that need to be represented as \emph{integer-weighted} graphs, usually referred to as \emph{multi-edge networks} or \emph{multi-graphs}.
The solution to this issue is to \emph{threshold} the interactions to obtain an unweighted graph (e.g.,~\citep{cranmer2011inferential}).
Clearly, this approach does not exploit all the information available in the data and therefore, may produce sub-optimal results~\citep{Ahnert2007}.

Addressing these limitations, ERGMs, for example, have been extended for count data~\citep{Krivitsky2017,Krivitsky2012}.
Furthermore, the latent space framework~\citep{Hoff2002} introduced a regression model that naturally admits different covariates and deals with count data as well~\citep{Sewell2015}.
The relational events model~\citep{Butts2008}, instead, handles interactions recorded along with a time stamp. 
This framework has since been extended to include, e.g., missing observations, auto-regressive models, and information to capture hidden homophily.
Other models that partially address those issues include the random dot product graphs~\citep{Sussman2012}, and the stochastic block models~\citep{Rohe2011}.
Using the most general theory, none of these models requires the observed interaction to be binary -- they can be counts or continuous~\citep{HOFF2013}.
However, in their generalisation to count or continuous data, these models require the assumption of a specific distribution for the number of edges, which implicitly assumes a distinctive edge generating process, as discussed in \citep{Casiraghi2019block}.
Moreover, many of these models do not scale well to large datasets.
In particular, in the case of ERGMs, the increased size of the sample space makes impractical the numerical estimation of model parameters employing Monte-Carlo simulations.
As a result, it is challenging to fit large datasets of repeated interactions.

The generalised hypergeometric ensemble of random graphs (gHypEG) allows to address these limitations, providing a suitable model for the analysis of complex systems~\citep{Casiraghi2017,Casiraghi2018,Casiraghi2019block}.
GHypEGs join two characteristics that are essential for the study of multi-edge networks.
First, they are specifically tailored to the analysis of multi-graphs, allowing the easy interpretation of parameters.
Second, their underlying probability distribution can be stated in closed form, thus simplifying the study of datasets with a large number of repeated interactions.
We demonstrate the power of our approach and its performance with an example based on an empirical dataset consisting of more than 180 000 interactions.
The available data consist of an interaction network, built from recorded contact counts between high-school students, and of further information such as student's gender, class membership and topic, self-reported friendship relations, and Facebook connections.

\section{Methodology}
\label{sec:methods}
\subsection{Network representation}
\begin{figure}[hbt]
\centering
  \includegraphics[width=.5\textwidth]{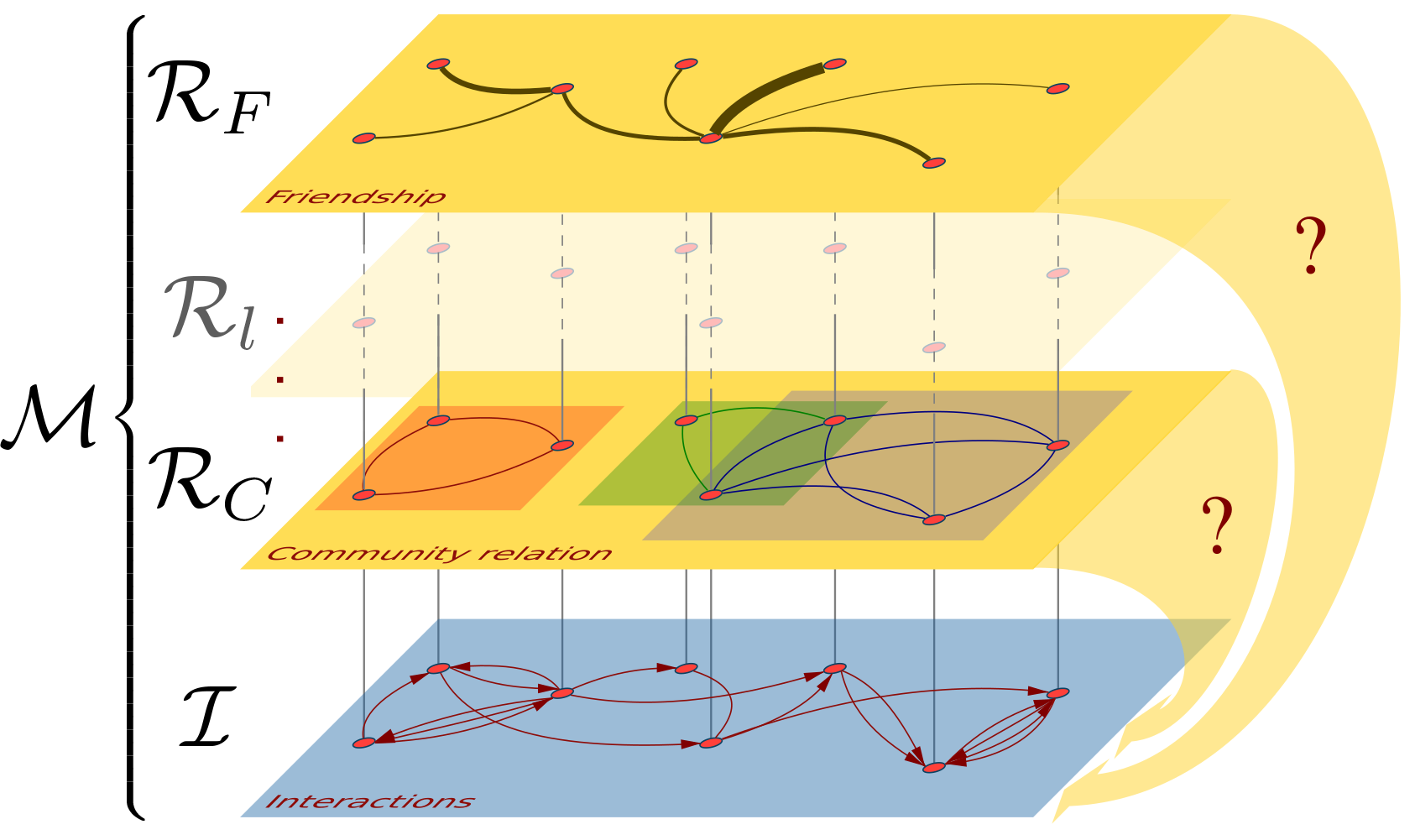}
  \caption[The multiplex network representation of a relational dataset.]{The multiplex network representation of a relational dataset. The bottom layer (blue) captures the interaction counts that are observed. The top layers (yellow) encodes different types of relations, like weighted friendship links, or community membership. The model we propose allows us to understand \emph{how} these relational layers impact interactions.}
  \label{fig:multiplex}
\end{figure}

Relational datasets as the one provided in \citep{Mastrandrea2015}, consist of interaction counts and a collection of dyadic relations and vertex attributes.
Vertex attributes, such as community membership or gender, often yield strong relations between individuals, as individuals in the same community tend to interact more than individuals in different ones.
We can study this type of data representing it as a \emph{multiplex network}.
Multiplex networks are a particular class of interconnected multi-layer networks where the vertices of each layer correspond (cf. \cref{fig:multiplex})~\citep{Garas2016}.

Suppose that we have a dataset consisting of $m$ recorded interactions between $n$ elements and $r$ different types of relations between them.
We encode the interactions in a graph with $n=\abs{V}$ vertices and $m$ (multi-)edges.
Since two individuals may interact more than once, multiple edges may exist between the same couple of vertices, giving rise to a \emph{multi-edge} graph.
In the following, we will refer to this graph as the interaction layer $\mathcal{I}$.
For each type of relation, we can generate a graph that encodes the dyadic relations between the elements of the system as \emph{weighted edges} between vertices.
The weight of each edge encodes the strength of the relation.
We will refer to these $r$ graphs as the relational layers $\mathcal R_l$ with $l\in[1,r]$.
Let now $\mathcal M$ be the multiplex network generated by the $r+1$ layers and $n=\abs{V}$ vertices.
\Cref{fig:multiplex} illustrates the multiplex approach we take.

In the following, we propose a framework to perform statistical regressions with these network layers as covariates.
We assume the multi-edged graph $\mathcal I$ to be the dependent variable and the remaining layers $\mathcal R_l$ to be the covariates, or explanatory variables.
The model that results has the following form:
\begin{equation}
  \label{eq:statmodel}
  \mathcal{I}=f(\mathcal R_1, \dots, \mathcal R_r; \theta_1, \dots, \theta_r),
\end{equation}
for some function $f:\mathbb R^{V\times V}\times \dots \times \mathbb R^{V\times V} \times \mathbb R^r\to \mathbb N^{V\times V}$, where the parameters $\theta_l$, $l\in [1,r]$ are the parameters of the regression model corresponding to each layer $\mathcal R_l$.

\subsection{Statistical Model}
\label{sec:fitting}
\paragraph{Generalised Hypergeometric Ensembles of Random Graphs (gHypEG)}\label{sec:ghype}

The approach described in this paper exploits the generalised hypergeometric ensemble of random graphs (gHypEG).
This class of models extends the configuration model (CM)~\citep{Molloy1995, molloy_reed_1998} by encoding complex topological patterns, while at the same time preserving degree distributions.
The aim of this article is to \emph{estimate} how to bias the process underlying the configuration model, based on observed data.
For this reason, before introducing the formulation of our regression model, we provide a brief overview of gHypEG.
A more formal presentation is given in~\citep{Casiraghi2016,Casiraghi2018,Casiraghi2019block}.

\begin{figure}[h!]
\centering
\includegraphics[width=.8\textwidth]{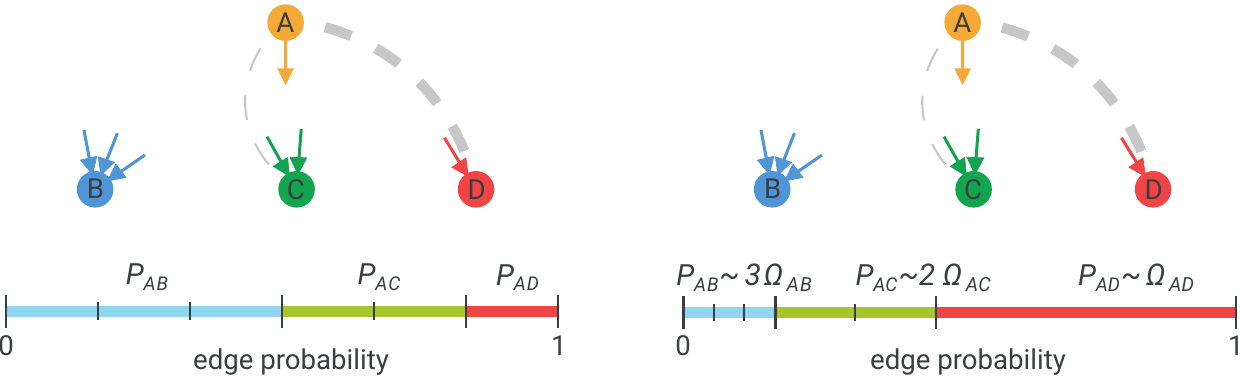}
\caption{Probabilities of connecting different stubs in CM and gHypEG.
Graphical illustration of the probability of connecting two vertices as a function of degrees (left figure), and degree and propensities (right figure).
Higher propensities can be related to strong relations between vertices, shown as dashed connection.
A stronger relation (thicker line) may result in a higher propensity to interact, as shown for the pair (A,D).}
\label{fig:props}
\end{figure}
In the CM, the probability of connecting two vertices depends only on their (out- and in-) degrees.
The CM assigns to each vertex as many out-stubs (or half-edges) as its out-degree, and as many in-stubs as its in-degree.
It then connects random pairs of vertices joining out- and in-stubs.
This is done by sampling uniformly at random one out- and one in-stub from the pool of all out- and in-stubs respectively, and then connecting them, until no more stubs are available~\citep{fosdick2018}.
The left side of \cref{fig:props} illustrates this case focusing on a vertex $A$.
The probability of connecting vertex $A$ with one of the vertices $B$, $C$, or $D$ depends only on the abundance of stubs, and hence on the in-degree of the vertices themselves.
The higher the in-degree, the higher the number of in-stubs of the vertex.
Hence, the higher the probability to randomly sample a stub belonging to the vertex.

GHypEG give an expression for the probability distribution underlying this process, where the degrees of the vertices are preserved in expectations~\citep{Casiraghi2018}.
This result is achieved exploiting an urn representation of the problem.
Edges are balls in an urn, and sampling from the CM corresponds to sampling balls (i.e., edges) from an urn constructed as follows.
For each pair of vertices $(i,j)$, we can denote with $k^{\text{out}}_i$ and $k^{\text{in}}_j$ their respective out- and in-degrees.
The number of combinations of out-stubs of $i$ with in-stubs of $j$ which could create an edge is given by $k^{\text{out}}_ik^{\text{in}}_j$.
For each dyad $(i,j)$ we place $k^{\text{out}}_ik^{\text{in}}_j$ balls of a given colour in the urn.
This provides us with an urn containing $\sum_{ij}k^{\text{out}}_ik^{\text{in}}_j$ edges of as many colours as pair of vertices that could be connected.
The process of sampling $m$ edges from such a `soft' configuration model is thus described by sampling $m$ balls from the urn, and the probability distribution of observing a graph $\G$ under the model is given by the \emph{multivariate hypergeometric distribution} with parameters $\bm\Xi=\{k^{\text{out}}_ik^{\text{in}}_j\}_{i,j}$:
\begin{equation}\label{eq:hyper}
    \Pr(\G\lvert\bm\Xi)=\dbinom{\sum_{ij}\Xi_{ij}}{m}^{-1}\prod_{i,j\in V}{\dbinom{\Xi_{ij}}{A_{ij}}},
\end{equation}
where $A_{ij}$ denotes the element $ij$ of the adjacency matrix of $\G$, and the probability of observing $\G$ is non-zero only if $\sum_{ij}A_{ij}=m$.

GHypEG expand this formulation to allow the modification of the CM based on observations about the system.
Specifically, we aim at modelling the probability of connecting two vertices not only based on degrees (i.e., number of stubs), but also on an independent \textbf{propensity} of two vertices to be connected.
Such propensities captures non-degree related effects to be incorporated into the model in the form of the odds of connecting a pair of vertices instead of another.
The right side of \cref{fig:props} illustrates this case, where $A$ is most likely to connect with vertex $D$, even though $D$ has only one available stub.

We can see this in the following way.
Suppose that there was an underlying social network connecting the vertices of $\G$.
Then, we could expect that vertices that have a strong connection in the social network (thick dashed line in \cref{fig:props}) have a high propensity to interact.
This results in a higher probability to observe interactions between the pair (A,D) compared to all others.
In \citep{Casiraghi2019block}, we have investigated how block and community structures can be encoded by specifying suitable propensities in the form of a block matrix.
Here, we look into the more general case where we aim at constraining the configuration model such that given edges are more likely than others according to external information about the process modelled.
Such external information will construct the covariates in our regression model.

We collect propensities in a matrix $\bm\Omega$.
The matrix encodes thus dyadic propensities of vertices that go beyond what prescribed by the combinatorial matrix $\bm\Xi$.
The ratio between any two elements $\Omega_{ij}$ and $\Omega_{kl}$ of the propensity matrix is the odds-ratio of observing an edge between vertices $i$ and $j$ instead of $k$ and $l$, independently of the degrees of the vertices.
The probability of a graph $\G$ depends on the stubs' configuration specified by $\bm\Xi$, and on the odds defined by $\bm\Omega$.
As for the case of the CM, this process can be seen as sampling edges from an urn, where edges characterised by a large propensity are more likely to be sampled.
Such a probability distribution is described by the multivariate Wallenius' noncentral hypergeometric distribution~\citep{wallenius1963, Chesson1978}:
\begin{equation}\label{eq:hyperI}
\Pr(\mathcal I|\bm\Xi,\bm\Omega)=\left[\prod_{i,j}{\dbinom{\Xi_{ij}}{A_{ij}}}\right]
         \int_{0}^{1}{\prod_{i,j}{\left(1-z^{\frac{\Omega_{ij}}{S_{{\Omega}} 
}}\right)^{A_{ij}}}dz}
\end{equation}
with $S_{{\Omega}}= \sum_{i,j} \Omega_{ij}(\Xi_{ij}-A_{ij})$.

Here, we assume that the entries of the matrix of stub's configuration $\bm\Xi$ are built according to the configuration model.
This is the most general way to encode combinatorial effect generated by the different activity, i.e., degree, of vertices.
It means that more active vertices, i.e., have a higher degree, are more likely to interact.
Hence, $\Xi$ is entirely defined by $\mathcal I$.

\paragraph{Regression Model}

The aim of our regression model is to find a suitable way to estimate $\bm\Omega$, based on the covariate layers $\{\mathcal R_l\}_{l\in[1,r]}$.
We thus propose to define $\bm\Omega$ as a function of the relational layers $\{\mathcal R_l\}_{l\in[1,r]}$:
\begin{equation}\label{eq:omega}
    \bm\Omega:=\prod_{l=1}^r{\left(\bm R^{(l)}\right)^{\theta_l}}=\exp\left\{\sum_{l=1}^r \theta_l \log \bm R^{(l)}\right\},
\end{equation}
where $\bm R^{(l)}$ is the adjacency matrix constructed from the network $\mathcal R_l$.
Under this assumption, we fix a multiplicative relation between the different layers.
That means, a value of $0$ for a dyad $i,j$ in any layer $\bm R^{(l)}$ corresponds to encoding the impossibility of observing any edge between $i$ and $j$.
Moreover, a value of $1$ for a dyad $i,j$ in a layer $\bm R^{(l)}$ means that the layer does not affect the probability of observing this dyad.
Furthermore, the right-hand side of \cref{eq:omega} provides a simple way to interpret the parameters of the model $\theta_l$.
If the covariate layers are specified in a convenient way, as we will show later, $\theta_l$ reflects the \textbf{log-odds} of observing an interaction between a pair of vertices for which there is an edge in the covariate layer $\mathcal R_l$, against a pair for which there is no edge in $\mathcal R_l$. 

We can now specify the statistical model in \cref{eq:statmodel}.
We take $f$ as the expectation of the gHypEG that maximises the probability of observing $\mathcal I$, given the relational layers $\{\mathcal R_l\}_{l\in[1,r]}$.
Estimating such a model is therefore equivalent to find maximum likelihood estimators (MLE) for the parameter vector $\bm\Theta$ in \cref{eq:hyper}.

\Cref{eq:hyper,eq:omega} show that the likelihood of $\bm\Theta$ given the observed graph $\mathcal I$ is defined by
\begin{equation}\label{eq:hyperlike}
L(\bm\Theta|\mathcal I)=\left[\prod_{i,j}{\dbinom{\Xi_{ij}}{A_{ij}}}\right]
         \int_{0}^{1}{\prod_{i,j}{\left(1-z^{\frac{\prod_{l=1}^r{\left(R^{(l)}_{ij}\right)^{\theta_l}}}{S_{\bm{\Theta}} 
}}\right)^{A_{ij}}}dz}
\end{equation}
with $S_{\bm{\Theta}}= \sum_{i,j} \prod_{l=1}^r{\left(R^{(l)}_{ij}\right)^{\theta_l}}(\Xi_{ij}-A_{ij})$.

Although the numerical maximisation of \cref{eq:hyperlike} is difficult, for $m\ll\sum_{ij}{\Xi_{ij}}$ we can approximate the Wallenius non-central multivariate hypergeometric distribution with a multinomial distribution with appropriately chosen probabilities $p_{ij}=\Xi_{ij}\Omega_{ij}/\sum_{kl}\Xi_{kl}\Omega_{kl}$ (cf.~\citep{zingg2019entropy}).
Because $\sum_{ij}{\Xi_{ij}}\approx m^2$, the multinomial approximation holds even for small networks.
Therefore \cref{eq:hyperlike} as a function of $\bm\Theta$ can be approximated up to constants by
\begin{equation}\label{eq:multinomlike}
    L(\bm\Theta|\mathcal I)\sim\prod_{i,j\in V}\left(\frac{\Xi_{ij}\prod_{l=1}^r \left(R^{(l)}_{ij}\right)^{\theta_l}}{\sum_{i,j\in V}\Xi_{ij}\prod_{l=1}^r \left(R^{(l)}_{ij}\right)^{\theta_l}}\right)^{A_{ij}}.
\end{equation}

We obtain the MLE $\bm{\hat\Theta}=\text{argmax}_\Theta(L(\bm\Theta|\mathcal I))$ of \cref{eq:multinomlike} by solving numerically the system given by $\nabla L(\bm\Theta)=0$.
Each component of the gradient of the log-likelihood $\nabla \log(L(\bm\Theta))$ is then given by
\begin{equation}
    \frac{\partial\log(L(\bm\Theta|\mathcal I))}{\partial \theta_l}=-m\frac{\sum_{ij}\log\left(R^{(l)}_{ij}\right)\Xi_{ij}\prod_{l=1}^r \left(R^{(l)}_{ij}\right)^{\theta_l}}{\sum_{ij}\Xi_{ij}\prod_{l=1}^r \left(R^{(l)}_{ij}\right)^{\theta_l}}+\sum_{ij}A_{ij}\log\left(R^{(l)}_{ij}\right)
\end{equation}

Thanks to the asymptotic properties of MLEs, we can compute the confidence intervals for the parameters estimates $\bm{\hat\Theta}$.
With $c$ as the appropriate $z$-critical value for a given confidence (e.g., $1.96$ for $95\%$ confidence intervals), the confidence interval for one parameter estimate $\hat\theta_l$ is given as follows:
\begin{equation}
    \label{eq:confint}
    \hat\theta_l \in \left[\hat\theta_l-c\sqrt{(\bm J(\bm{\hat\Theta})^{-1})_{ll}} , \hat\theta_l+c\sqrt{(\bm J(\bm{\hat\Theta})^{-1})_{ll}}\right],
\end{equation}
where $\bm J(\bm{\hat\Theta}) = -\nabla^2 \log(L(\bm{\hat\Theta}|I))$ is the observed Fisher information matrix~\citep{Degroot2002}.
From \cref{eq:multinomlike} we get the following expression for $\bm J(\bm{\hat\Theta})$:
\begin{equation}
\begin{aligned}
    \bm J(\bm{\hat\Theta})_{lk}=
    m\frac{
    \left[\sum_{ij}\Xi_{ij}\prod_{l=1}^r \left(R^{(l)}_{ij}\right)^{\theta_l}\right]\left[\sum_{ij}\log\left(R^{(l)}_{ij}\right)\log(R_{k,ij})\Xi_{ij}\prod_{l=1}^r \left(R^{(l)}_{ij}\right)^{\theta_l}\right]
    }{
    \left[\sum_{ij}\Xi_{ij}\prod_{l=1}^r \left(R^{(l)}_{ij}\right)^{\theta_l}\right]^2
    }+\\
    -m\frac{
    \left[\sum_{ij}\log\left(R^{(l)}_{ij}\right)\Xi_{ij}\prod_{l=1}^r \left(R^{(l)}_{ij}\right)^{\theta_l}\right]\left[\sum_{ij}\log(R_{k,ij})\Xi_{ij}\prod_{l=1}^r \left(R^{(l)}_{ij}\right)^{\theta_k}\right]
    }{
    \left[\sum_{ij}\Xi_{ij}\prod_{l=1}^r \left(R^{(l)}_{ij}\right)^{\theta_l}\right]^2
    }
\end{aligned}
\end{equation}

In the \texttt{R} library \texttt{ghypernet}~\citep{ghypernet}, available to download from the CRAN, we provide the \texttt{nrm} routine to perform network regression model estimation.

\subsection{General regression model}
\label{sec:regression}

The model described in the previous section can be generalized to account for multiple observations of the multiplex $\mathcal M$.
For example, suppose we have data about contacts between students in a school, and we have collected the same type of data for different schools.
Let us assume now we want to learn whether gender homophily plays the same role in the interactions across all the schools.
This implies that while the relations between the individuals change for different observations, e.g., gender distribution in different schools, the effect that the relations have on the interactions remains constant.
In other words, the relational layers change for each observation, i.e., $\mathcal R^{(i)}\neq\mathcal R^{(j)}$, where $i$ and $j$ are different observations.
On the other hand, the parameter $\theta$ quantifying the effect of the relations on the interactions is assumed to be constant, i.e., $\theta^{(i)}=\theta^{(j)}=\theta$.

Suppose thus to have $N$ independent observations of the multiplex $\mathcal M$,  each denoted $\mathcal M^{(i)}$.
We assume that the influence of the independent layers $R_l^{(i)}$ on the dependent layer $\mathcal I^{(i)}$ is fixed, i.e., for each observation $i$, $\theta^{(i)}=\theta\,\forall i\in N$. 

Since each observation $\mathcal I^{(i)}$ is independent and follows the distribution of the gHypEG given in \cref{eq:hyper}, the joint probability distribution is just the product of each probability.
Therefore the likelihood of the parameter vector $\bm\Theta$ is given by
\begin{equation}\label{eq:jointlike}
    L(\bm\Theta|\mathcal I^{(0)}, \mathcal I^{(1)}, \dots, \mathcal I^{(N)}):=\prod_{i=1}^N L(\bm\Theta|\mathcal I^{(i)}),
\end{equation}
where $L(\bm\Theta|\mathcal I^{(i)})$ is defined as in \cref{eq:hyperlike}.
It is worth noting that the interaction layers $\mathcal I^{(i)}$ come from the same class of distributions but are not identically distributed.
This is true unless the number of edges $M^{(i)}=M$ and the matrix $\Xi^{(i)}=\Xi$ are constant for each observation $(i)$.

Given the likelihood in \cref{eq:jointlike}, we can derive the MLE $\bm{\hat\Theta}$ of the parameter $\theta$.
Denoting with $L(\bm\Theta|\mathcal I^{(i)})$ the log-likelihood of $\theta$ and by
\begin{equation}
    \bar L(\bm\Theta)=\frac{1}{N}\sum_{i=1}^N L(\bm\Theta|\mathcal I^{(i)})
\end{equation}
the average log-likelihood, $\bm{\hat\Theta}$ is defined as follows:
\begin{equation}
    \bm{\hat\Theta}=\text{argmax}_\theta\left(\bar L(\bm\Theta)\right).    
\end{equation}

\subsection{Model selection and effect sizes}\label{sec:test}
Recall we have a multiplex $\mathcal M$ with $r+1$ layers.
Suppose we have estimated the statistical regression model defined in \cref{sec:regression}.
We thus know the MLEs $\{\hat\theta_l\}_{l\in[1,r]}$ corresponding the $r$ relational layers $\{\mathcal R_l\}_{l\in[1,r]}$, and each of their values quantifies the \emph{strength} of the effect each layer has on the interaction layer $\mathcal I$.
Are all these parameters needed?
In other words, we want to quantify the goodness of fit of the model with all parameters $\{\hat\theta_l\}_{l\in[1,r]}$, and compare it to a model with fewer parameters.
This allows us to select the parameters and the layers with significant effect, and disregard those with non-significant effects on the interactions.

We want to compare which of two statistical models defined by the sub-multiplexes $\{\mathcal R_l\}_{l\in[1,q]}$ and $\{\mathcal R_l\}_{l\in[1,q+s]}$ as in \cref{eq:statmodel}, one with $q$ and the other one with $q+s$ relational layers, better describes the observed interaction layer $\mathcal I$.

Both models are described by \cref{eq:hyperlike} with the appropriate layers chosen as predictors.
The two models are nested, as one is a particular case of the other.
In fact, the model defined by $\{\mathcal R_l\}_{l\in[1,q]}$ can be obtained by setting to $0$ the $s$ coefficients $\{\mathcal \theta_l\}_{l\in[q,q+s]}$ corresponding to the $\{\mathcal R_l\}_{l\in[q,q+s]}$ layers in the second model (cf. \cref{eq:omega}).

We can perform \emph{model selection} using the \emph{likelihood ratio test}.
In particular, we can identify the null hypothesis $H_0$ by the model defined by $\{\mathcal R_l\}_{l\in[1,q]}$ with $\tilde q$ parameters, and the alternative hypothesis $H_1$ by the model defined by $\{\mathcal R_l\}_{l\in[1,q+s]}$, with $\tilde s$ more parameters.
This allows testing whether the explaining power of the more complex model with $\tilde q + \tilde s$ parameters is high enough to justify the increase in complexity.

Alternatively, as discussed already in \citep{Casiraghi2019block}, we can use AIC and BIC to choose the best between the two models.
Moreover, information criteria allow us to compare \emph{all} models built using different combinations of layers, even when they are not nested.
If we proceed in a step-wise fashion, constructing the sub-multiplexes corresponding to the set of predictors with decreasing AIC scores, we obtain a forward selection method that allows building models of increasing complexity.

Finally, the goodness of fit of the model can be assessed qualitatively through the adjusted McFadden's pseudo-r-squared $\rho^2$~\citep{mcfadden1973}.
The adjusted McFadden's pseudo-r-squared is a coefficient of determination analogous to the multiple-correlation coefficient used in OLS linear regression models, adjusted for model complexity.
It is based on maximum likelihood estimates of model parameters, and it is hence suitable to evaluate the goodness of fit of our model.
It is defined as follows:
\begin{equation}\label{eq:rho2}
  \rho^2 = 1 - \frac{L(\bm{\hat\Theta}_{q+s}|\mathcal I)-K}{L(\bm\Theta_0|\mathcal I)}.
\end{equation}
In \cref{eq:rho2}, $L(\bm{\hat\Theta}|\mathcal I)$ is the log-likelihood of the full model obtained from the MLE $\bm{\hat\Theta}_{q+s}$ of the parameter vector $\theta_{q+s}$, $L(\bm\Theta_0|\mathcal I)$ is the likelihood of the null-model where no explanatory variable is used (i.e., the CM), and $K$ is number of degrees of freedom of the full model.
The closer the value of $\rho^2$ is to 1, the better is the fit of the model.
The inclusion of the number of degrees of freedom $K$ adjusts for model complexity, by punishing models with an excessive number of parameters.
The value of $\rho^2$ can also be seen as an estimate of the amount of variability in the data explained by the model, in terms of the relative increase in likelihood of the model.

More generally, we can define a McFadden coefficient
\begin{equation}\label{eq:mc}
  \text{MC} = 1 - \frac{L(\bm{\hat\Theta}_{q+s}|\mathcal I)}{L(\bm{\hat\Theta}_{q}|\mathcal I)}.
\end{equation}
that allows to evaluate the relative increment in likelihood provided by extending the model defined by $\{\mathcal R_l\}_{l\in[1,q]}$ with $q$ parameters with $s$ more parameters.
Large values of the McFadden coefficient can then be used to proxy the improvement generated by the addition of new predictors which are introducing new information into the model.
Low values of the McFadden coefficient, on the other hand, reflect the introduction of predictors that do not improve the model considerably and thus do not allow to obtain new insights on the data.

In the next section, we show an application of these methods on an empirical dataset about human interactions.
\section{Application: High School Contacts Analysis}
\label{sec:applications}
\subsection{Data}
We showcase our method with a case study.
Specifically, we apply our technique to a SocioPattern dataset~\citep{Mastrandrea2015}, to measure the strength and the significance of the effect of each layer of information provided on the observed number of interactions.
At \url{https://www.sg.ethz.ch/nrm-tutorial}, we provide a tutorial companion to this article with the code used to generate the results shown here.

In this case study, we analyse a dataset consisting of 188 508 recorded contacts between $327$ students over five days that we represent in the graph of interactions $\G$.
The dataset contains additional types of relations between the students that we encode as predictors in different relational layers.
The available relations, that serve as covariates for the regression model are the following.
There are 2 social networks providing connection between the students.
One social network contains self-reported (directed) friendship relations.
The second social network reports Facebook connection between students, which result in undirected links.
Furthermore, students are assigned to 9 different classes, grouped into 4 topical blocks.
We hypothesise that students in the same class and in the same topical block are more likely to interact with each other.
Moreover, the gender of each student is provided.
These last 3 layers are thus built according to categorical information about the vertices.
In practice, these require defining a block model for each category and then estimate them all together.
However, here, because we want to be able to join together categorical data with dyadic data, we proceed differently than in \citep{Casiraghi2019block}.

\begin{figure}[hbt]
\centering
  \includegraphics[width=.6\textwidth]{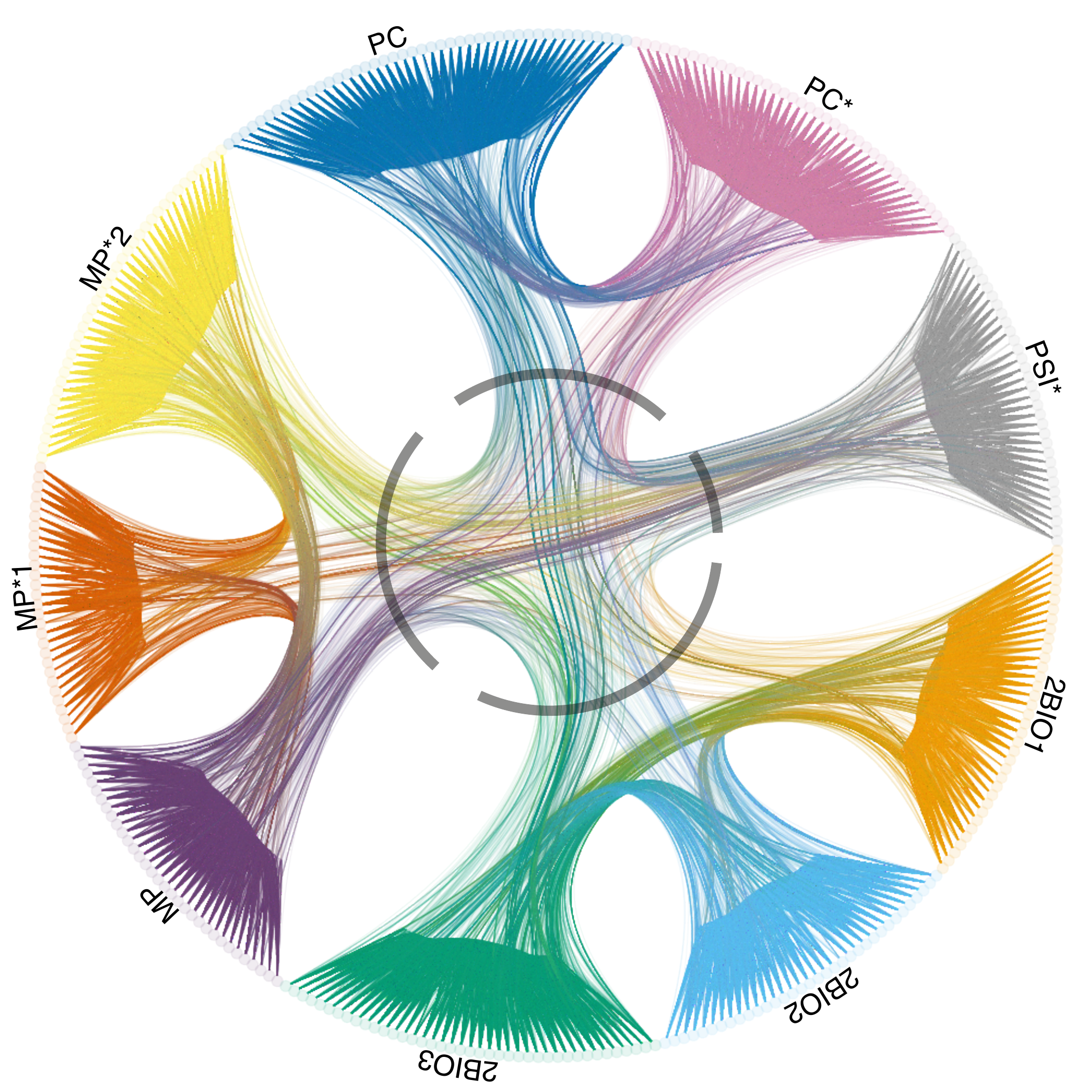}
  \caption[The graph obtained from the contacts between students.]{The graph obtained from the contacts between students. Each student is coloured according to its class membership and the internal ring groups classes on a similar topic. From this figure, it is clear that most of the contacts happen between students of the same class, and there is a preference for contacts between students attending classes on the same topic.}\label{fig:graph}
\end{figure}

Let $l$ be a labelling of vertices and $\mathcal R_l$ the corresponding layer in the multiplex representation.
We can define a partition vector $z^l$, whose i-th entry $z^l_i$ specifies the label of vertex $i$.
For every dyad $i,j$ for which $z^l_i=z^l_j$, we set $R^{(l)}_{ij}=\kappa$.
In the cases where $z^l_i\neq z^l_j$, we set $R^{(l)}_{ij}=1$.
When performing the MLE of the parameter $\theta_l$, corresponding to the layer $\mathcal R_l$, we rescale the value of $R^{(l)}$ such that $(R^{(l)})^{\theta_l}$ estimates the strength of the effect provided by the labelling $l$.
Note that we could choose any value for $\kappa$, as what we are interested into is the MLE propensity $\kappa^\theta$, which will be constant, satisfying \cref{eq:hyperlike}.
If we fix $\kappa=e$, where $e$ is Euler's number, $\theta_l$ can be easily interpreted in terms of the order of magnitude of the contribution of $R^{(l)}$ to the propensity matrix of the gHypEG.
Furthermore, if we define the odds-ratio $\omega=\Omega_{ij}/\Omega_{kl}$, we can see that the log-odds $\log(\omega)$ are given by $\theta$:
\begin{equation}
  \log(\omega)=\log\left(\frac{\Omega_{ij}}{\Omega_{kl}}\right)=\log\left[\left(\frac{R^{(l)}_{ij}}{R^{(l)}_{kl}}\right)^{\theta_l}\right]=\theta_l\log\left(\frac{R^{(l)}_{ij}}{R^{(l)}_{kl}}\right)=\theta_l\log\left(\frac{e}{1}\right)=\theta_l
\end{equation}

If $\theta_l$ is larger than $0$, there is a positive effect, as $\kappa^\theta_l>1$ is \emph{increasing} the propensity $\Omega_{ij}$ for $i,j$ with the same label.
Similarly, if $\theta_l<0$ there is a negative effect, i.e., the graph is disassortative with respect to the labelling $l$.

The first relational layer $\mathcal R_C$ in the dataset reflects the separation of students into 9 classes.
We want to control for the separation into classes, as the encounters between students attending different classes are naturally limited, as can be observed in \cref{fig:graph}.
To build $\bm R^{(C)}$ we can set $R^{(C)}_{ij}=e$ if $i,j$ are in the same class, and $R^{(C)}_{ij}=1$ if $i,j$ are in different classes.

A second relational layer $\mathcal R_T$ is built according to the topic of the different classes students take.
The nine classes are grouped into four topical areas, of 3,3,2,1 classes, respectively.
There are three classes of type "MP" (MP, MP01, MP02), two of type "PC" (PC and PC0), one of type "PSI" (PSI0) and 3 of type "BIO" (2BIO1, 2BIO2, 2BIO3).
This separation is highlighted by the internal ring in \cref{fig:graph}.
The layer $\mathcal R_T$ is defined similarly to $\mathcal R_C$, setting $R^{(T)}_{ij}=e$ if $i,j$ attend classes in the same topical area, and $R^{(T)}_{ij}=1$ if not.

The third relational layer $\mathcal R_G$ is built using the gender of the students.
We want to correct for gender homophily, as this could partially play a role in student interactions.
We build its adjacency matrix $\bm R^{(G)}$ as above.

The dataset also provides information about actual friendship relations between the students.
We can build the fourth predictor using the social networks obtained from \emph{self-reported friendship relations}.
Because this data is self-reported, it generates a directed social network.
In fact, some students report a friendship relation with another student, which have no corresponding link from the other student.
For this reason, we can model this predictor as \textbf{two separate layers}, given that the interactions on which we want to regress are undirected.
We set one layer $\mathcal R_{f}$ to capture all corresponded friendships, i.e., all those edges that are symmetric.
We set a second layer $\mathcal R_{1/2f}$ to capture all non-corresponded friendships, i.e., all those edges that are asymmetric.
Both layers are built following a similar process to the one discussed above: if there is a friendship relation between two vertices $i,j$, we set the value of the adjacency matrix of the corresponding layer to $R^{(f)}_{ij}=\kappa=e$.
Otherwise, we set the value in the adjacency matrix to $1$.
This way, we can interpret the parameter $\theta_f$ as the log-odds of observing an interaction between two `friends' against two non-`friends'.

The fifth predictor is built using the provided \emph{Facebook connections}.
This dataset is, according to \citep{Mastrandrea2015}, incomplete.
In fact, not all students disclosed their Facebook accounts to extract relations.
That means, for some students, we know the presence or the absence of Facebook relationships, while for others, we cannot say anything.
Instead of modelling the lack of information as a lack of relations, we can split this predictor into two non-separable layers.
We do so by building a layer $\mathcal R_\text{fb}$ similarly to the friendship layer.
In particular, we set $R^{(\text{fb})}_{ij}=1$ for all those dyads for which we have no data.
Moreover, we generate a dummy `correction' layer $\mathcal R_{\varepsilon}$ where we set $R^{(\varepsilon)}_{ij}=e$ for all those dyads for which we have no data, and $R^{(\varepsilon)}_{ij}=1$ otherwise.
In this way, we can estimate the true effect of the available Facebook relations, correcting for the effect of the non-available ones.
Because in this case
\begin{equation}
  \log(\omega^\text{friend}_\text{non friend})=
  \log\left(\frac{\Omega_\text{fb friend}}{\Omega_\text{non-friend}}\right)=
  \log\left[\left(\frac{R^{(\text{fb})}_\text{fb friend}}{R^{(\text{fb})}_\text{non friend}}\right)^{\theta_\text{fb}}\cdot
  			\left(\frac{R^{(\varepsilon)}_\text{data}}{R^{(\varepsilon)}_\text{data}}\right)^{\theta_\varepsilon}\right]=
  \theta_\text{fb},
\end{equation}
\begin{equation}
  \log(\omega^\text{friend}_\text{no data})=
  \log\left(\frac{\Omega_\text{fb friend}}{\Omega_\text{no data}}\right)=
  \log\left[\left(\frac{R^{(\text{fb})}_\text{fb friend}}{R^{(\text{fb})}_\text{no data}}\right)^{\theta_\text{fb}}\cdot
  			\left(\frac{R^{(\varepsilon)}_\text{data}}{R^{(\varepsilon)}_\text{no data}}\right)^{\theta_\varepsilon}\right]=
  \theta_\text{fb}-\theta_\varepsilon,
\end{equation}
$\theta_\text{fb}$ provides the log-odds of interactions between students that are friends on Facebook against those between students that are not friends.
Similarly, $\theta_\text{fb}-\theta_\varepsilon$ gives the log-odds of interactions between students that are friends on Facebook against those with a student that did not provide access to the data.

In general, we assume that the absence of an edge in either $\mathcal R_f$ or $\mathcal R_\text{fb}$ is not enough to disallow an interaction to happen.
It is for this reason that we choose to set the weight of the relations between students who are not "friends" in either of the two layers to $1$.
If we were to set it to $0$ instead, we would have disallowed the presence of edges between those dyads in the model entirely, in contrast with what observed in the data.

We speculate that $\mathcal R_C$ will have a very strong influence on the interactions since the division into classes acts as physical boundary for students interactions.
In general, moreover, we would assume the information provided by the two friendship layers will be comparable, as the reported friendship relations should be part of the Facebook connections.
Similarly, we expect that corresponded friendship will yield a stronger effect on interactions.

\subsection{Model}
We build a regression model with the five predictors described above.
With such a model, we answer the question of whether interactions between students are related to (a) friendship relations, as perceived by the students themselves, and (b) Facebook connections.
The estimated effects are corrected for the degree of the vertices, i.e., for how active students are, and for the the fact that the students are phisically separated in different classes.
Hence, as a first step we estimate a model for the case (a) and the case (b).
The first two columns of \cref{tab:beta} provide the estimates of $\bm{\Theta^{(a)}}$ and $\bm{\Theta^{(b)}}$ respectively.
In both cases, we see that there is a strong effect provided by the two social networks, signalled by a positive value of the estimated parameters.
Also, we see that the offline social network defined by the friendship relations has as a stronger effect compared to the online social network.
This can be seen both from the effect size highlighted by the absolute value of the parameters, and from the larger value of $\rho^2$ and smaller AIC.
The second two columns in \cref{tab:beta} show the model estimated after correcting for the control variables defined by $\mathcal R_C$, $\mathcal R_T$, $\mathcal R_G$.
We notice that, while the effect of friendship relations remains strong, the effect of Facebook connections almost entirely disappears when controlling for the class membership of students.
Finally, in the full model shown in the fifth column of \cref{tab:beta} it can be seen that friendship relations completely take over the small explaining power produced by Facebook connections.
\begin{table}
\caption[Fitted parameters for the 7-layer model, standard errors for the estimates, and corresponding significance of the parameter estimates, obtained from a standard t-test as described in \cref{eq:confint}.]{Fitted parameters for the 7-layer model, standard errors for the estimates, and corresponding significance of the parameter estimates, obtained from a standard t-test as described in \cref{eq:confint}. 3 stars correspond to a p-value $p<\alpha=0.001$. For the regression, we used $\kappa=e$.}\label{tab:beta}
\footnotesize
\begin{center}
\begin{tabular}{l D{.}{.}{7.6} D{.}{.}{7.6} D{.}{.}{6.6} D{.}{.}{6.6} D{.}{.}{6.6} }
\toprule
 & \multicolumn{1}{c}{$\bm\Theta^{(a)}$} & \multicolumn{1}{c}{$\bm\Theta^{(b)}$} & \multicolumn{1}{c}{$\bm\Theta^{(a\dagger)}$} & \multicolumn{1}{c}{$\bm\Theta^{(b\dagger)}$} & \multicolumn{1}{c}{$\bm\Theta^{(a\dagger+b\dagger)}$} \\
\midrule
Control                       &             &             &             &             &             \\
                              &             &             &             &             &             \\
\quad $\bm R^{(C)}$           &             &             & 3.196^{***} & 3.318^{***} & 3.168^{***} \\
                              &             &             & (0.011)     & (0.011)     & (0.011)     \\
\quad $\bm R^{(T)}$           &             &             & 2.275^{***} & 2.281^{***} & 2.281^{***} \\
                              &             &             & (0.021)     & (0.021)     & (0.021)     \\
\quad $\bm R^{(G)}$           &             &             & 0.194^{***} & 0.258^{***} & 0.200^{***} \\
                              &             &             & (0.005)     & (0.005)     & (0.005)     \\
Friendship      				  &             &             &             &             &             \\
                              &             &             &             &             &             \\
\quad $\bm R^{(f)}$           & 3.696^{***} &             & 1.810^{***} &             & 1.820^{***} \\
                              & (0.005)     &             & (0.006)     &             & (0.006)     \\
\quad $\bm R^{(1/2f)}$        & 2.147^{***} &             & 0.385^{***} &             & 0.421^{***} \\
                              & (0.015)     &             & (0.015)     &             & (0.015)     \\
Facebook                      &             &             &             &             &             \\
                              &             &             &             &             &             \\
\quad $\bm R^{(\text{fb})}$   &             & 2.344^{***} &             & 0.535^{***} & 0.106^{***} \\
                              &             & (0.006)     &             & (0.006)     & (0.006)     \\
\quad $\bm R^{(\varepsilon)}$ &             & 0.564^{***} &             & 0.330^{***} & 0.357^{***} \\
                              &             & (0.005)     &             & (0.005)     & (0.005)     \\
\midrule\midrule
AIC                      	  & 516342.8    & 643916.5    & 4061.8      & 71444.8     & 0           \\
$\rho^2$			          & 0.175       & 0.081       & 0.556       & 0.506       & 0.559       \\
\bottomrule
\multicolumn{6}{l}{\scriptsize{$^{***}p<0.001$, $^{**}p<0.01$, $^*p<0.05$}}
\end{tabular}
\end{center}
\end{table}

As we expected, from the results of the regression we can see a strong effect obtained from the separation of vertices into the categories corresponding to classes.
In the full model, the value of $\theta_C\gg0$ implies an odds-ratio of $e^{\theta_C}=23.76$ for the probability of an interaction between classmates against an encounter of students of different classes, given everything else equal.
This means that there are approximately $24$ more chances that two classmates meet, compared to encounters between students of different classes.
Class topics are also a driving force for the interactions.
Contact between students attending classes on the same topic is $10$ times more likely to be observed than contact between students attending classes on different topics. 
The value of $\theta_G$ supports the presence of a weak gender homophily in the encounters between students, with an odds-ratio of $1.22$.
The effect of self-reported friendship is large and positive as expected, while non-corresponded friendships yield a much lower effect, even though this is larger than that provided by Facebook connections.

We now proceed to study the contribution of each relational layer to model fit.
To do so, we follow a stepwise selection method, as described in \cref{sec:test}.
We introduce one predictor after the other, starting from those that have the highest contribution according to AIC.
This means that, in the first step, we add the predictor whose corresponding model has the lowest AIC.
Then, in the second we add one at a time to the first predictor all the remaining ones, to find the second-best contribution, and we proceed until all predictors have been added.

\begin{table}
\caption[Model selection steps.]{Model selection steps.
         For each step, we report the McFadden R squared, the improvement in AIC, and the relative goodness-of-fit in terms of the MC coefficient. The 6 models are ordered by increasing complexity.
         }\label{tab:aic}
\footnotesize
\begin{center}
\begin{tabular}{l D{.}{.}{5.6} D{.}{.}{5.6} D{.}{.}{4.6} D{.}{.}{4.6} D{.}{.}{3.6} D{.}{.}{2.6} }
\toprule
 & \multicolumn{1}{c}{(1)} & \multicolumn{1}{c}{(2)} & \multicolumn{1}{c}{(3)} & \multicolumn{1}{c}{(4)} & \multicolumn{1}{c}{(5)} & \multicolumn{1}{c}{(6)} \\
\midrule
Control                       &             &             &             &             &             &             \\
                              &             &             &             &             &             &             \\
\quad $\bm R^{(C)}$           & 4.641^{***} & 4.417^{***} & 3.207^{***} & 3.179^{***} & 3.176^{***} & 3.168^{***} \\
                              & (0.007)     & (0.007)     & (0.008)     & (0.008)     & (0.008)     & (0.008)     \\
\quad $\bm R^{(T)}$           &             &             & 2.307^{***} & 2.314^{***} & 2.282^{***} & 2.281^{***} \\
                              &             &             & (0.016)     & (0.016)     & (0.016)     & (0.016)     \\
\quad $\bm R^{(G)}$           &             &             &             &             & 0.205^{***} & 0.200^{***} \\
                              &             &             &             &             & (0.004)     & (0.004)     \\
Friendship      &             &             &             &             &             &             \\
                              &             &             &             &             &             &             \\
\quad $\bm R^{(f)}$           &             & 1.819^{***} & 1.812^{***} & 1.817^{***} & 1.801^{***} & 1.820^{***} \\
                              &             & (0.004)     & (0.004)     & (0.004)     & (0.004)     & (0.005)     \\
\quad $\bm R^{(1/2f)}$        &             &             &             &             &             & 0.421^{***} \\
                              &             &             &             &             &             & (0.011)     \\
Facebook                      &             &             &             &             &             &             \\
                              &             &             &             &             &             &             \\
\quad $\bm R^{(\text{fb})}$   &             &             &             & 0.121^{***} & 0.124^{***} & 0.106^{***} \\
                              &             &             &             & (0.005)     & (0.005)     & (0.005)     \\
\quad $\bm R^{(\varepsilon)}$ &             &             &             & 0.347^{***} & 0.351^{***} & 0.357^{***} \\
                              &             &             &             & (0.004)     & (0.004)     & (0.004)     \\
\midrule\midrule
AIC                           & 98609.8   & 21323.0   & 6370.6    & 2514.3    & 713.7     & 0.0       \\
MC                        & 0.486 					& 0.112 & 0.024 & 0.006 & 0.003 & 0.001\\
$\rho^2$         & 0.486       & 0.543       & 0.554       & 0.557       & 0.558       & 0.559       \\
\bottomrule
\multicolumn{7}{l}{\scriptsize{$^{***}p<0.001$, $^{**}p<0.01$, $^*p<0.05$}}
\end{tabular}
\end{center}
\end{table}

During the stepwise selection process, we monitor the increment in goodness-of-fit in terms of AIC, and the relative goodness-of-fit in terms of the McFadden coefficient MC.
As described above, these two criteria give us two alternative ways to perform model selection.
By looking at the change in AIC (cf. \cref{tab:aic}), we see a clear difference between the first two models and the remaining ones.
In fact, both predictors $\mathcal R_C$ and $\mathcal R_f$ provide a substantial decrease in AIC.
In other words, they have good explaining power for the observed interactions.
On the other hand, the other predictors provide a less marked decrease in AIC.
We also observe that the second two models, corresponding to the introduction of the layers $\mathcal R_T$ (topic) and the couple $\mathcal R_\text{fb}$ and $\mathcal R_\varepsilon$ (fb), provide a similar decrease in AIC.
Finally, the last two predictors provide a smaller decrease in AIC.
In terms of information, however, the best model according to AIC is nevertheless the one that incorporates all parameters~\citep{Lehmann3}.

If we consider the relative improvement in likelihood instead, as provid\-ed by the McFadden coefficient, we see a similar pattern.
The first two parameters provide a definite improvement in the goodness of fit, the second two a small improvement, while the last two show a negligible improvement.
The reason for these results has to be searched in the fact that the predictors are partially correlated.
In fact, the class predictor and the friendship predictors provide largely independent data.
The third predictor, although important, is a superset of the class predictor.
Hence, it yields a smaller improvement in the goodness of fit of the model.
A solution to this issue could be obtained by modelling the different classes as separated blocks in a BCCM (cf.~\citep{Casiraghi2019block}, increasing the number of parameters but capturing both the class and the topic membership at the same time.
The fourth predictor, reporting Facebook relations, is incomplete.
Hence, it can only explain part of the data.
Moreover, it is partly correlated with self-declared friendship, as people that declare to be friends are often friends on Facebook (40\% of the two social networks overlap).

From this example, we can conclude that in the dataset studied the observed interactions are strongly influenced by social relations in the form of friendship links, even when we correct for the subdivision into classes and topic, as can also be visualised in \cref{fig:graph}.
Gender homophily is, instead, relatively weak after accounting for all factors.
Moreover, it is interesting to note that non-corresponded friendships, i.e., friendships that have been declared only by one student, have a very low effect on the observed interactions, as long as corresponded friendships are taken into account.

\section{Conclusion}
\label{sec:conclusion}

In this article, we have proposed a new statistical model to quantify how observed interactions depend on different relations, in the framework of multiplex networks.
The model is based on the assumption that interactions between elements of a system are driven by two factors.
The first factor is the existence of relations between elements, such as friendship or homophily.
The second is the combinatorial randomness caused by the activity of the elements.
Elements that are more active are more likely to interact with each other, even if they are unrelated.

Different from common approaches used in network analysis, our methodology has been specifically designed to deal with  multi-edge graphs.
It therefore allows to use the whole data available, without the need of thresholding it to obtain unweighted, i.e., binary, graphs.
In fact, repeated interactions between elements of a system generate multi-edge graphs, where the vertices correspond to the elements of the system.
Similarly, relations can have varying intensity and should be encoded in weighted graphs.
This is why thresholding the data into binary networks can be a waste of useful information.

Our model separates random and deterministic influences on interactions, accounting for the randomness as combinatorial effects.
We hence identify \emph{how much} known relations drive the interactions.
To achieve this, we base our regression model on generalized hypergeometric ensembles of random graphs, a class of statistical network ensembles we have recently introduced.
The formulation of our model allows to estimate the strength of the dependence between relations and interactions, together with its statistical significance.
Moreover, the parameters estimated by our model can be readily interpreted as the log-odds of observing interactions between two different pairs of vertices.
Thanks to these characteristics, the statistical regression model described in this article results in a powerful tool for the analysis of complex systems consisting of a large number of highly interacting elements.

Studying how different relations drive observed interactions is not only necessary to increase the understanding of a system, it is also needed to control the dynamics of a system.
In fact, to do so we have to appropriately modify the relations that are the driving forces underlying its behavior.
Similarly, if we want to increase the resilience of a system, we want to affect the relations that are responsible for its weaknesses.
Having a clear understanding on how and which relations impact the behavior of the elements of a system is a necessary condition to properly control it.

In conclusion, the method we propose is a major advance for the analysis of relational datasets and complex networks.
By allowing the study of multi-edge and weighted graphs, it increases the breadth of applicability of network theory.
In future work, it will allow to identify missing interactions, according to null-models based on known relations.
Thanks to this, it will be possible to uncover \emph{unknown relations} between elements of a system.

\section*{Acknowledgment}
The author thanks S. Schweighofer, G. Vaccario and F. Schweitzer for useful discussion, and V. Nanumyan for designing \cref{fig:graph}.

\end{document}